\documentstyle[11pt,psfig,twoside]{article}

\begin{document}

\newcommand{\dd}{\,{\rm d}}
\newcommand{\ie}{{\it i.e.},\,}
\newcommand{\etal}{{\it et al.\ }}
\newcommand{\eg}{{\it e.g.},\,}
\newcommand{\cf}{{\it cf.\ }}
\newcommand{\vs}{{\it vs.\ }}
\newcommand{\zdot}{\makebox[0pt][l]{.}}
\newcommand{\up}[1]{\ifmmode^{\rm #1}\else$^{\rm #1}$\fi}
\newcommand{\dn}[1]{\ifmmode_{\rm #1}\else$_{\rm #1}$\fi}
\newcommand{\upd}{\up{d}}
\newcommand{\uph}{\up{h}}
\newcommand{\upm}{\up{m}}
\newcommand{\ups}{\up{s}}
\newcommand{\arcd}{\ifmmode^{\circ}\else$^{\circ}$\fi}
\newcommand{\arcm}{\ifmmode{'}\else$'$\fi}
\newcommand{\arcs}{\ifmmode{''}\else$''$\fi}
\newcommand{\MS}{{\rm M}\ifmmode_{\odot}\else$_{\odot}$\fi}
\newcommand{\RS}{{\rm R}\ifmmode_{\odot}\else$_{\odot}$\fi}
\newcommand{\LS}{{\rm L}\ifmmode_{\odot}\else$_{\odot}$\fi}

\newcommand{\Abstract}[2]{{\footnotesize\begin{center}ABSTRACT\end{center}
\vspace{1mm}\par#1\par
\noindent
{~}{\it #2}}}

\newcommand{\TabCap}[2]{\begin{center}\parbox[t]{#1}{\begin{center}
  \small {\spaceskip 2pt plus 1pt minus 1pt T a b l e}
  \refstepcounter{table}\thetable \\[2mm]
  \footnotesize #2 \end{center}}\end{center}}

\newcommand{\TableSep}[2]{\begin{table}[p]\vspace{#1}
\TabCap{#2}\end{table}}

\newcommand{\FigCap}[1]{\footnotesize\par\noindent Fig.\  %
  \refstepcounter{figure}\thefigure. #1\par}

\newcommand{\TableFont}{\footnotesize}
\newcommand{\TableFontIt}{\ttit}
\newcommand{\SetTableFont}[1]{\renewcommand{\TableFont}{#1}}

\newcommand{\MakeTable}[4]{\begin{table}[htb]\TabCap{#2}{#3}
  \begin{center} \TableFont \begin{tabular}{#1} #4 
  \end{tabular}\end{center}\end{table}}

\newcommand{\MakeTableSep}[4]{\begin{table}[p]\TabCap{#2}{#3}
  \begin{center} \TableFont \begin{tabular}{#1} #4 
  \end{tabular}\end{center}\end{table}}

\newenvironment{references}%
{
\footnotesize \frenchspacing
\renewcommand{\thesection}{}
\renewcommand{\in}{{\rm in }}
\renewcommand{\AA}{Astron.\ Astrophys.}
\newcommand{\AAS}{Astron.~Astrophys.~Suppl.~Ser.}
\newcommand{\ApJ}{Astrophys.\ J.}
\newcommand{\ApJS}{Astrophys.\ J.~Suppl.~Ser.}
\newcommand{\ApJL}{Astrophys.\ J.~Letters}
\newcommand{\AJ}{Astron.\ J.}
\newcommand{\IBVS}{IBVS}
\newcommand{\PASP}{P.A.S.P.}
\newcommand{\Acta}{Acta Astron.}
\newcommand{\MNRAS}{MNRAS}
\renewcommand{\and}{{\rm and }}
\section{{\rm REFERENCES}}
\sloppy \hyphenpenalty10000
\begin{list}{}{\leftmargin1cm\listparindent-1cm
\itemindent\listparindent\parsep0pt\itemsep0pt}}%
{\end{list}\vspace{2mm}}

\def\TYLDA{~}
\newlength{\DW}
\settowidth{\DW}{0}
\newcommand{\dw}{\hspace{\DW}}

\newcommand{\refitem}[5]{\item[]{#1} #2%
\def\REFARG{#3}\ifx\REFARG\TYLDA\else, {\it#3}\fi
\def\REFARG{#4}\ifx\REFARG\TYLDA\else, {\bf#4}\fi
\def\REFARG{#5}\ifx\REFARG\TYLDA\else, {#5}\fi.}

\newcommand{\Section}[1]{\section{#1}}
\newcommand{\Subsection}[1]{\subsection{#1}}
\newcommand{\Acknow}[1]{\par\vspace{5mm}{\bf Acknowledgements.} #1}
\pagestyle{myheadings}

\def\thefootnote{\fnsymbol{footnote}}
\begin{center}
{\large\bf The Optical Gravitational Lensing Experiment.\\
%\vskip3pt
Cepheids in the Magellanic Clouds.\\
%\vskip3pt
V.  Catalog of Cepheids from the Small Magellanic Cloud\footnote
{Based on  observations obtained with the 1.3~m Warsaw telescope at the
Las Campanas  Observatory of the Carnegie Institution of Washington.}}
\vskip0.8cm
{\bf
A.~~U~d~a~l~s~k~i$^1$,~~I.~~S~o~s~z~y~{\'n}~s~k~i$^1$,
~~M.~~S~z~y~m~a~{\'n}~s~k~i$^1$,
~~M.~~K~u~b~i~a~k$^1$,~~G.~~P~i~e~t~r~z~y~\'n~s~k~i$^1$,
~~P.~~W~o~\'z~n~i~a~k$^2$,~~ and~~K.~~\.Z~e~b~r~u~\'n$^1$}
\vskip3mm
{$^1$Warsaw University Observatory, Al.~Ujazdowskie~4, 00-478~Warszawa, Poland\\
e-mail: (udalski,soszynsk,msz,mk,pietrzyn,zebrun)@astrouw.edu.pl\\
$^2$ Princeton University Observatory, Princeton, NJ 08544-1001, USA\\
e-mail: wozniak@astro.princeton.edu}
\end{center}

\Abstract{
We present the Catalog of Cepheids from the SMC which contains data for
2049 objects detected in the 2.4 square degree area of central parts of
the SMC. For each object  period, {\it BVI} photometry, astrometry, and
$R_{21}, \phi_{21}$ parameters of the Fourier decomposition of {\it
I}-band light curve are provided. The Catalog is based on observations
collected during the OGLE-II microlensing survey.

Tests of completeness performed in overlapping parts of adjacent fields
indicate that completeness of the Catalog is very high: $\approx92$\%.
Statistics and distributions of basic parameters of Cepheids are also
presented. 

All presented data, including individual {\it BVI} observations
($\approx4.7\cdot10^5$ {\it BVI} measurements), are available from the
OGLE Internet archive.

}{~}

\Section{Introduction}

Cepheids belong to one of the most important astrophysical objects.
Those pulsating variable stars are well known from their famous
Period--Luminosity $(P-L)$  relation, discovered at the beginning of the
20th century (Leavitt 1912), making these objects widely recognized
standard candle for distance determination. Observations of Cepheids
also provide important tests on stellar evolution, stellar structure
etc.

Unfortunately, the observational data of Cepheids collected during
the past decades and available for testing properties of these important
objects were highly inhomogeneous: collected by many astronomers with
different instruments etc. The situation dramatically changed  in
1990s when the large microlensing searches began regular photometric
monitoring of the Magellanic Clouds. Precise CCD photometry of millions
of stars in these galaxies is a natural by-product of microlensing
surveys and the Magellanic Clouds are known to possess large population
of Cepheids. Both the MACHO and EROS microlensing projects presented
results of observations of Cepheids in the Magellanic Clouds providing
new  interesting information on these stars (Alcock \etal 1995, Alcock
\etal 1999, Sasselov \etal 1997, Bauer \etal 1999). Unfortunately, all
these data were taken in non-standard photometric bands.

The Magellanic Clouds were also included to the list of targets of the
Optical Gravitational Lensing Experiment microlensing search at the
beginning of the second phase of the project (OGLE-II) in January 1997.
After two years of constant photometric monitoring of both galaxies with
the standard {\it BVI} filters closely resembling the standard system,
the collected observational material is large enough so the search for
variable stars, in particular for Cepheids, could be performed.

The main goal of the series of papers on Cepheids in the Magellanic
Clouds is to provide large, homogeneous and statistically complete
samples of Cepheids from the Magellanic Clouds -- high quality data for
testing properties of Cepheids. In the previous papers of the series we
presented large sample of double mode Cepheids in the SMC (Udalski \etal
1999a), first candidates for single mode second overtone Cepheids
(Udalski \etal 1999b), analysis of the $P-L$ and $P-L-C$ relations based
on large samples of LMC and SMC Cepheids (Udalski \etal 1999c) and the
Catalog of Cepheids in the LMC consisting of about 1300 objects (Udalski
\etal 1999d). In this paper we continue the series presenting the
largest sample of Cepheids from one environment -- the Catalog of
Cepheids from the SMC consisting of about 2150 objects. 

Both Catalogs constitute an ideal observational data set for many
projects concerning Cepheids. Similarly to the Catalog of Cepheids from
the LMC, all data presented in this paper including individual
observations ($\approx 4.7\cdot10^5$ measurements) are available to the
astronomical community from the OGLE Internet archive.

\Section{Observations}

All observations presented in this paper were carried out during the
second  phase of the OGLE experiment with the 1.3-m Warsaw telescope at
the Las  Campanas Observatory, Chile, which is operated by the Carnegie
Institution of  Washington. The telescope was equipped with the "first
generation" camera with a SITe ${2048\times2048}$ CCD detector working
in drift-scan mode. The pixel size was 24~$\mu$m giving the 0.417
arcsec/pixel scale. Observations of  the SMC were performed in the
"slow" reading mode of CCD detector with the  gain 3.8~e$^-$/ADU and
readout noise of about 5.4~e$^-$. Details of the  instrumentation setup
can be found in Udalski, Kubiak and Szyma{\'n}ski  (1997). 

\MakeTable{lcc}{12.5cm}{Equatorial coordinates of the SMC fields}
{
\hline
\noalign{\vskip3pt}
\multicolumn{1}{c}{Field} & RA (J2000)  & DEC (J2000)\\
\hline
\noalign{\vskip3pt}
SMC$\_$SC1  & 0\uph37\upm51\ups & $-73\arcd29\arcm40\arcs$\\
SMC$\_$SC2  & 0\uph40\upm53\ups & $-73\arcd17\arcm30\arcs$\\
SMC$\_$SC3  & 0\uph43\upm58\ups & $-73\arcd12\arcm30\arcs$\\
SMC$\_$SC4  & 0\uph46\upm59\ups & $-73\arcd07\arcm30\arcs$\\
SMC$\_$SC5  & 0\uph50\upm01\ups & $-73\arcd08\arcm45\arcs$\\
SMC$\_$SC6  & 0\uph53\upm01\ups & $-72\arcd58\arcm40\arcs$\\
SMC$\_$SC7  & 0\uph56\upm00\ups & $-72\arcd53\arcm35\arcs$\\
SMC$\_$SC8  & 0\uph58\upm58\ups & $-72\arcd39\arcm30\arcs$\\
SMC$\_$SC9  & 1\uph01\upm55\ups & $-72\arcd32\arcm35\arcs$\\
SMC$\_$SC10 & 1\uph04\upm51\ups & $-72\arcd24\arcm45\arcs$\\
SMC$\_$SC11 & 1\uph07\upm45\ups & $-72\arcd39\arcm30\arcs$\\  
\hline}

Regular observations of the SMC started on June~26, 1997. For two
fields, SMC$\_$SC5 and SMC$\_$SC6, several {\it VI}-band frames were
also collected in January~1997. 11 driftscan fields  covering
$14.2\times 57$~arcmins on the sky were observed covering in total about
2.4~square degrees. The microlensing search  is planned to last for
several years, thus observations of selected fields will be  continued
during the following seasons.  In this paper we present data  collected
up to March 1999.

Observations were obtained in the standard {\it  BVI}-bands. The
effective exposure time was 125, 174 and 237 seconds for the {\it I, V}
and {\it B}-band, respectively. Due to microlensing search observing
strategy the vast majority of observations were done through the {\it
I}-band filter (about $120-200$ epochs depending on the field) while
images on about $15-40$ epochs were collected in the {\it BV}-bands. The
instrumental system closely resembles the standard {\it BVI} one -- the
color coefficients of transformation ($a\cdot CI$; $a$ -- color
coefficient, $CI$ -- color index: $B-V$ for {\it B} and $V-I$ for {\it
VI} filters) are equal to $-0.041$, $-0.002$ and $+0.029$ for the {\it
B, V} and {\it I}-band, respectively.  Collected images were reduced
with the standard OGLE data pipeline. Quality of the photometric data of
the SMC is described in Udalski \etal (1998b). In particular, the
accuracy of absolute photometry zero points is about $0.01-0.02$~mag in
all {\it BVI}-bands.

\begin{figure}[htb]
\vspace*{11cm}
\FigCap{OGLE-II fields in the SMC. Dots indicate positions of Cepheids
from the Catalog. North is up and East to the left in this Digitized Sky
Survey image of the SMC.}
\end{figure}

Table~1 lists equatorial coordinates of the center of each field and its
acronym. Fig.~1 shows the Digitized Sky Survey image of the SMC with
contours of the observed fields.

\Section{Selection of Cepheids}

Selection of Cepheid candidates was performed in the identical way as
for the LMC fields (Udalski \etal 1999d). In short, all stars with
unusually large standard deviation as compared to non-variable stars of
similar brightness were subject to period search procedure based on AoV
alghoritm (Schwarzenberg-Czerny 1989).  Typically about 120--200 epochs
were available for each analyzed  object with the lower limit set to~50.
The mean {\it I}-band magnitude of analyzed objects was limited to
${I<20.0}$~mag. 

Candidates for Cepheids were selected from the entire sample of variable
stars  based on visual inspection of their light curves and location in
the color-magnitude diagram (CMD) within the area limited by
${I{<}18.5}$~mag and  ${0.25{<}(V{-}I){<}1.3}$~mag. Several objects
located outside this region (\eg highly  reddened Cepheids) and objects
with no color information but with evident Cepheid-type light curves
were also included to this sample. In total about 2280 Cepheid
candidates were found in the searched area of the  SMC center. 

Each of the analyzed SMC fields overlaps with neighboring fields for 
calibration purposes. Therefore several dozen Cepheids located in the 
overlapping regions were detected twice. Similarly to the LMC Cepheid
catalog we decided not to remove them from the final list of objects
because their measurements are independent in both fields and can be
used for testing quality of data, completeness of the sample etc. 118
such objects were detected and we provide cross-reference list to
identify them.

\Section{Basic Parameters of Candidates}

\Subsection{Intensity Mean Photometry}

{\it BVI} intensity mean photometry of each object from our sample of
Cepheid candidates  was derived by integrating the light curve converted
to intensity units. The light curve  was  approximated by the Fourier
series of fifth order and results were converted back to the magnitude
scale. Statistical accuracy of the mean {\it I}-band photometry is about
$0.001-0.005$~mag and somewhat worse (about 0.01~mag) for poorer sampled
{\it BV}-bands \ie much smaller than uncertainty of zero points of
standard photometry.

For each object we also determined the extinction insensitive index
$W_I$ (called also Wesenheit index, Madore and Freedman 1991):

$$ W_I=I-1.55*(V-I) \eqno{(1)}  $$

The coefficient 1.55 in Eq.~(1) corresponds to the coefficient resulting
from standard interstellar extinction curve dependence of the {\it
I}-band extinction on $E(V-I)$ reddening (\eg Schlegel, Finkbeiner and
Davis 1998). It is easy to show that the values of $W_I$  are the same
when derived from observed or extinction free magnitudes, provided that
extinction to the object is not too high so it can be approximated with
a linear function of color.

\Subsection{Interstellar Reddening}

Determination of the interstellar reddening  to the SMC Cepheids was
performed in similar way as for the LMC objects (Udalski \etal 1999d).
We used red clump stars for mapping the fluctuations of mean reddening
in our observed fields treating their mean {\it I}-band magnitude as the
reference brightness. It was shown to be independent on age of these
stars in the wide range of $2-10$~Gyr, and it is only slightly dependent
on metallicity (Udalski 1998a,b).  Thus, the mean brightness of red
clump stars can be a very good reference of brightness for monitoring
extinction. Similar method was used by Stanek (1996) for determination
of the extinction map of Baade's Window in the Galactic bulge.

The reddening in the SMC is smaller and more homogeneous than in the LMC
what can be assessed from the shape of red clump in subsequent fields.
Therefore we only determined reddening in 11 lines-of-sight -- one per
entire OGLE field. In each line-of-sight we determined the mean observed
{\it I}-band magnitude of red clump stars (Table~2) with technique
identical to that described in Udalski \etal (1998a). Differences of the
observed {\it I}-band magnitudes were assumed as differences of the mean
$A_I$ extinction. We converted differences of $A_I$ extinction to
differences of $E(B-V)$ reddening assuming the standard extinction
curve: $E(B-V)=A_I/1.96$ (Schlegel \etal 1998).

\MakeTable{lcc}{12.5cm}{$E(B-V)$ reddening in the SMC fields.}
{\hline
\multicolumn{1}{c}{Field}& $\langle I_{\rm RC} \rangle$ & $E(B-V)$\\
\hline
SMC$\_$SC1 & 18.457 & 0.070\\
SMC$\_$SC2 & 18.473 & 0.078\\
SMC$\_$SC3 & 18.494 & 0.089\\
SMC$\_$SC4 & 18.505 & 0.094\\
SMC$\_$SC5 & 18.518 & 0.101\\
SMC$\_$SC6 & 18.504 & 0.094\\
SMC$\_$SC7 & 18.510 & 0.097\\
SMC$\_$SC8 & 18.517 & 0.100\\
SMC$\_$SC9 & 18.469 & 0.076\\
SMC$\_$SC10& 18.475 & 0.079\\
SMC$\_$SC11& 18.485 & 0.084\\
\hline}

The zero points of our reddening map were derived based on previous
determinations in two lines-of-sight -- toward star clusters NGC416
(Mighell, Sarajedini and French 1998) and NGC330 (Caloi \etal 1993). The
former determination is based on analysis of photometry of the cluster
while the latter on IUE observations of OB stars. Both these zero points
were consistent with our map to within a few thousandths of magnitude.

We also checked the absolute calibration of our map comparing the
observed {\it I}-band magnitude of red clump stars with extinction free
magnitude determined from a few star clusters in the halo of the SMC
(Udalski 1998b). The calibration {\it via} extinction free magnitude of
red clump stars gave slightly smaller (by about 0.01~mag) zero point of
the $E(B-V)$ reddening. Therefore we conservatively adopted the error of
our map as equal  to $\pm0.02$~mag.  The final $E(B-V)$ reddening in the
SMC is listed in Table~2. Interstellar extinction in the {\it BVI} bands
was calculated using the standard extinction curve coefficients (\eg
Schlegel \etal 1998):

$$ A_B=4.32\cdot E(B-V) $$
$$ A_V=3.24\cdot E(B-V) $$
$$ A_I=1.96\cdot E(B-V) $$

\Subsection{Astrometry} 

Equatorial coordinates of all candidates were calculated based on
transformation derived with the Digitized Sky Survey  images. Details of
procedure are described in Udalski \etal (1998b). About $3000-8000$
stars common in OGLE and DSS images (depending on stellar density of the
field) were used for transformation. Internal accuracy of the equatorial
coordinates is about 0.15 arcsec with possible systematic errors of the
DSS coordinate system up to 0.7 arcsec.

\Subsection{Fourier Parameters of Light Curve Decomposition}

Shape of the light curve is often used for analyses of  properties of
pulsating stars and discrimination between pulsating modes. Therefore we
derived Fourier parameters $R_{21}=A_2/A_1$ and
$\phi_{21}=\phi_2-2\phi_1$ of the Fourier series decomposition of {\it
I}-band light curve. $A_i$ and $\phi_i$ are the amplitudes and phases of
$(i-1)$  harmonic of the Fourier decomposition of light curve.
Parameters $R_{21}$ and $\phi_{21}$ are commonly  used for analyses of
the shape of light curves.

Fifth order Fourier series were fitted to the magnitude scale {\it
I}-band light curve. In the case of objects with almost sinusoidal light
curve for which the first harmonic amplitude and phase were not
statistically significant, $R_{21}=0$ and $\phi_{21}$ is not defined. 

\begin{figure}[htb]
\hglue-0.5cm\psfig{figure=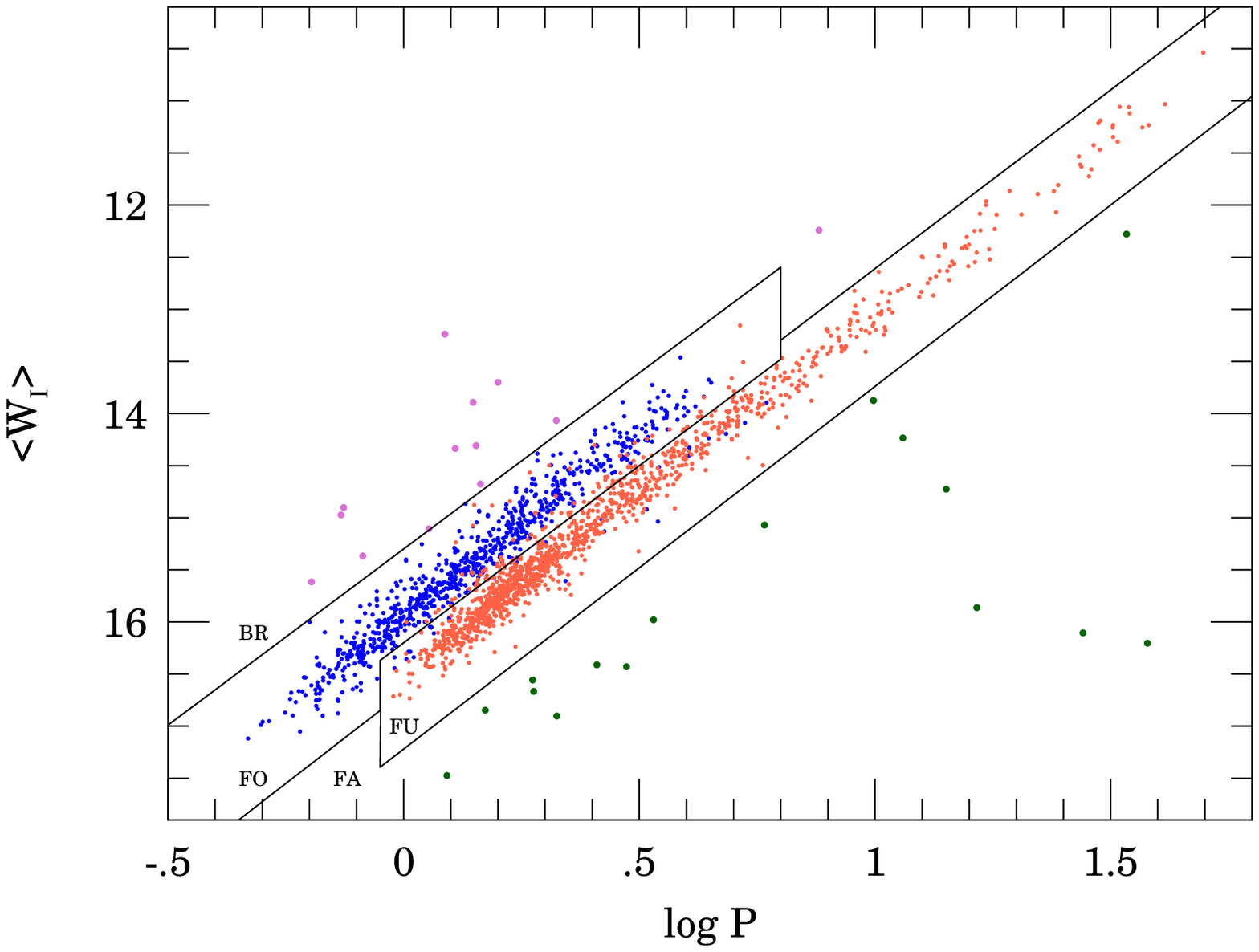,bbllx=30pt,bblly=50pt,bburx=510pt,bbury=410pt,width=12.5cm,clip=}
\vspace*{3pt}
\FigCap{Period-Luminosity relation for extinction insensitive index
$W_I$. Contours divide the diagram into sections where  in the first
approximation fundamental (FU) and first overtone mode (FO) classical
Cepheids are found. Section denoted by BR indicates region where objects
were classified as brighter than FO Cepheids and by FA -- as fainter
than FU Cepheids. Small dots mark positions of objects finally
classified as FU and FO classical Cepheids (light and dark dots,
respectively). Larger dots -- BR (light dots) and FA (dark dots)
objects.}
\end{figure}

\Subsection{Classification}

Classification of objects from our sample of Cepheid candidates was
performed in two steps. In the first approximation we divided all
objects into four groups: classical Cepheids pulsating in the
fundamental mode (FU), classical Cepheids pulsating in the first
overtone mode (FO), objects brighter than FO mode Cepheids (BR) and
objects fainter than FU mode Cepheids (FA).  The division was made based
on the $P-L$ diagram constructed for the extinction insensitive index
$W_I$. Fig.~2 presents $P-L$ diagram for the $W_I$ index with boundaries
of these four regions.

Unfortunately, due to much larger geometrical depth of the SMC as
compared to the LMC the separation between the FU and FO Cepheids in the
$W_I$ $P-L$ diagram is not that sharp and sequences of both types of
pulsating stars overlap. Therefore we made the final classification
after inspection of location of each FU and FO Cepheid in the $R_{21}$
and $\phi_{21}$ \vs $\log P$ diagrams. It is well known that such
diagrams allow to separate between the FU and FO mode pulsators (\cf
Alcock \etal 1999, Udalski \etal 1999a). Sequences for FU and FO
Cepheids in $R_{21}$ \vs $\log P$ diagram are well separated (except for
$0.6<\log P < 0.8$) and in most cases classification is straightforward.
Separation between  FU and FO Cepheids in $\phi_{21}$ \vs $\log P$
diagram is much worse but it still can be used for classification in
some period ranges. Many objects which were preliminarly classified as FU
or FO objects based on  the $P-L$ diagram were shifted to the opposite
group after the second test.

Fig.~3 presents the final $R_{21}$ \vs $\log P$ and $\phi_{21}$ \vs
$\log P$ diagrams for all objects classified as FU and FO mode classical
Cepheids. Objects finally classified as FU and FO Cepheids are shown
with different type of dots (FU -- light dots, FO -- dark dots) in
Fig.~2.

\begin{figure}[p]
\psfig{figure=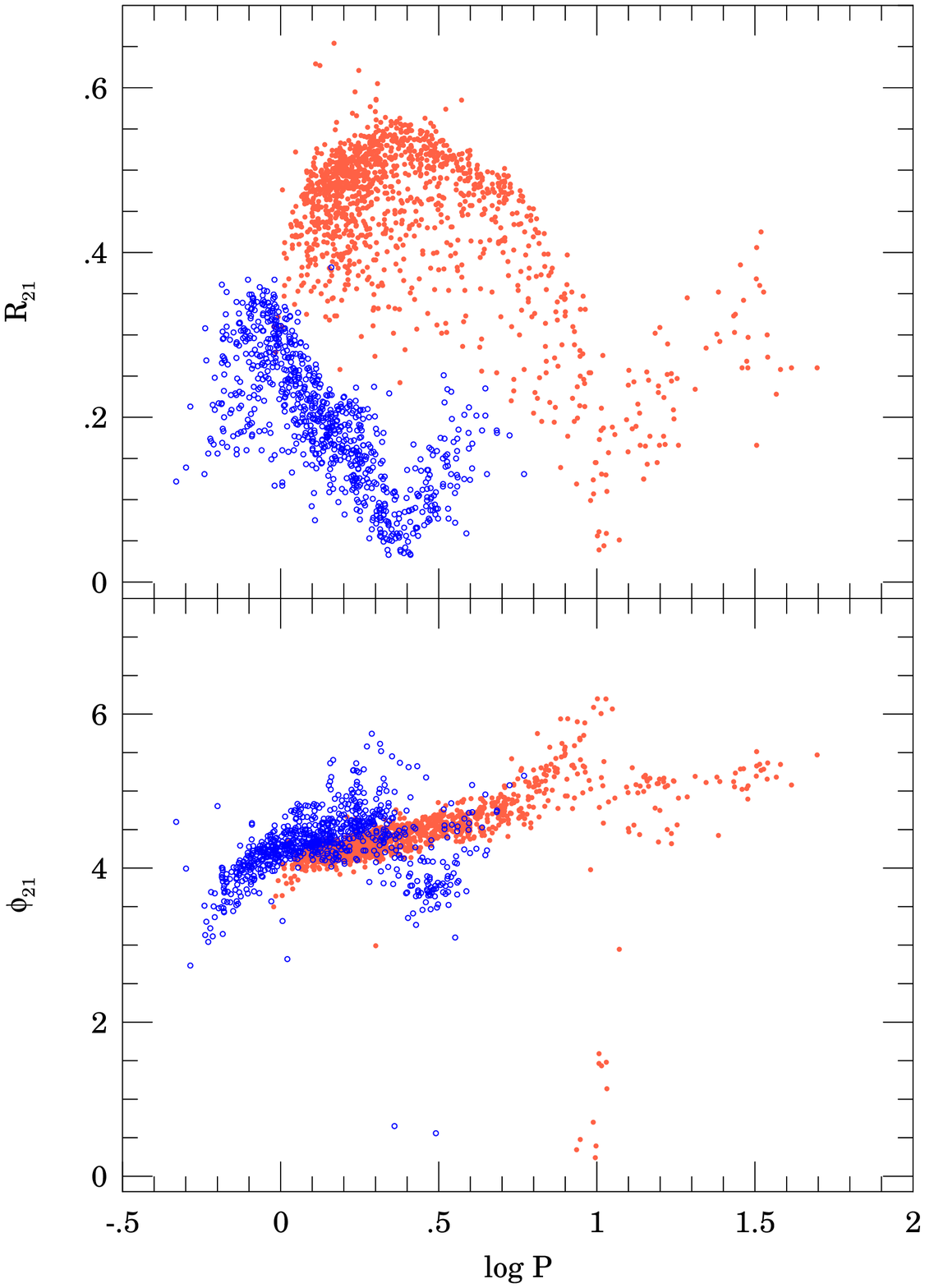,bbllx=30pt,bblly=40pt,bburx=505pt,bbury=705pt,width=12.5cm
,clip=}
\FigCap{${R_{21}}$ and ${\phi_{21}}$ \vs $\log P$ diagrams for
single-mode classical Cepheids from the SMC. Dark open circles indicate
positions of the first overtone Cepheids while light dots positions of
fundamental mode pulsators.}
\end{figure}

\Section{Catalog of Cepheids from the SMC}

2167 Cepheid candidates passed our selection criteria. They are listed
in Table~3. First column of Table~3 is the star identification: {\it
field$\_$name star$\_$num\-ber}. In the next columns the equatorial
coordinates, RA and DEC (J2000), period in days and moment of the zero
phase corresponding to maximum light are given followed by intensity
mean {\it IVB} photometry and  extinction insensitive index $W_I$. In
the next two columns Fourier parameters, $R_{21}$ and $\phi_{21}$, of
the light curve decomposition are listed. Finally, in the last column
classification of the object is provided.

Table~3 contains 2167 entries but only 2049 objects: 118 stars were
detected twice -- in the overlapping regions of adjacent fields. Table~4
provides cross-identification of all such objects. With additional
samples of double-mode and second overtone Cepheids from the SMC
(Udalski \etal 1999a,b), the total number of Cepheids discovered in the
OGLE-II SMC fields is equal to 2155. 

The {\it I}-band light curves of objects from Table~3 are presented in
Appendices A--L. The ordinate is phase with 0.0 value corresponding to
maximum light. Abscissa is the {\it I}-band  magnitude. The light curve
is repeated twice for clarity.

Finding charts ($60\arcs\times 60\arcs$ part of the {\it I}-band image)
are not presented in this paper but they are available in electronic
form from the OGLE Internet archive (see below).

Similarly to the LMC Cepheid catalog we did not attempt to
cross-identify our objects with the ones known from literature.  Because
of high completeness of the Catalog  (see Section~6)  and precise {\it
BVI} photometry the OGLE catalog is likely to supersede much of the
previous works. If necessary, cross-identification with selected objects
can be done with precise coordinates and finding charts provided with
the Catalog.

One should also remember about the limit of the Catalog on the brighter,
\ie longer period object side because of saturation of the CCD detector.
This limit corresponds to objects with period longer than $\log
P\approx1.7$, \ie longer than about 50~days.

\renewcommand{\TableFont}{\scriptsize}
%\vspace*{-25pt}
\setcounter{table}{3}
\MakeTableSep{
l@{\hspace{3pt}}
r@{\hspace{4pt}}
c@{\hspace{4pt}}
l@{\hspace{3pt}}
r@{\hspace{30pt}}
l@{\hspace{3pt}}
r@{\hspace{4pt}}
c@{\hspace{4pt}}
l@{\hspace{3pt}}
r}{12.5cm}{Cross-identification of stars detected in overlapping regions}
{
SMC$\_$SC1  & 93903 & $\leftrightarrow$ & SMC$\_$SC2 &	    33 &  SMC$\_$SC5  &311512 & $\leftrightarrow$ & SMC$\_$SC6 &   67302 \\
SMC$\_$SC1  & 95571 & $\leftrightarrow$ & SMC$\_$SC2 &	  1423 &  SMC$\_$SC5  &311542 & $\leftrightarrow$ & SMC$\_$SC6 &   67255 \\
SMC$\_$SC1  &101220 & $\leftrightarrow$ & SMC$\_$SC2 &	  5856 &  SMC$\_$SC5  &311544 & $\leftrightarrow$ & SMC$\_$SC6 &   67265 \\
SMC$\_$SC2  & 81006 & $\leftrightarrow$ & SMC$\_$SC3 &	    31 &  SMC$\_$SC5  &311598 & $\leftrightarrow$ & SMC$\_$SC6 &   72589 \\
SMC$\_$SC2  & 86445 & $\leftrightarrow$ & SMC$\_$SC3 &	  8759 &  SMC$\_$SC6  &263571 & $\leftrightarrow$ & SMC$\_$SC7 &    4030 \\
SMC$\_$SC2  & 90387 & $\leftrightarrow$ & SMC$\_$SC3 &	 15993 &  SMC$\_$SC6  &263687 & $\leftrightarrow$ & SMC$\_$SC7 &    4123 \\
SMC$\_$SC2  &101512 & $\leftrightarrow$ & SMC$\_$SC3 &	 35855 &  SMC$\_$SC6  &267872 & $\leftrightarrow$ & SMC$\_$SC7 &    3976 \\
SMC$\_$SC2  &107212 & $\leftrightarrow$ & SMC$\_$SC3 &	 43000 &  SMC$\_$SC6  &267903 & $\leftrightarrow$ & SMC$\_$SC7 &    4007 \\
SMC$\_$SC3  &178269 & $\leftrightarrow$ & SMC$\_$SC4 &      62 &  SMC$\_$SC6  &267952 & $\leftrightarrow$ & SMC$\_$SC7 &    4041 \\
SMC$\_$SC3  &189138 & $\leftrightarrow$ & SMC$\_$SC4 &	  5209 &  SMC$\_$SC6  &272425 & $\leftrightarrow$ & SMC$\_$SC7 &    8731 \\
SMC$\_$SC3  &189247 & $\leftrightarrow$ & SMC$\_$SC4 &	  5277 &  SMC$\_$SC6  &276930 & $\leftrightarrow$ & SMC$\_$SC7 &   13526 \\
SMC$\_$SC3  &198055 & $\leftrightarrow$ & SMC$\_$SC4 &	 11348 &  SMC$\_$SC6  &281567 & $\leftrightarrow$ & SMC$\_$SC7 &   22908 \\
SMC$\_$SC3  &202810 & $\leftrightarrow$ & SMC$\_$SC4 &	 18836 &  SMC$\_$SC6  &296772 & $\leftrightarrow$ & SMC$\_$SC7 &   37298 \\
SMC$\_$SC3  &202971 & $\leftrightarrow$ & SMC$\_$SC4 &	 14944 &  SMC$\_$SC6  &296773 & $\leftrightarrow$ & SMC$\_$SC7 &   37299 \\
SMC$\_$SC3  &208364 & $\leftrightarrow$ & SMC$\_$SC4 &	 18792 &  SMC$\_$SC6  &306480 & $\leftrightarrow$ & SMC$\_$SC7 &   47121 \\
SMC$\_$SC3  &208591 & $\leftrightarrow$ & SMC$\_$SC4 &	 19028 &  SMC$\_$SC6  &306527 & $\leftrightarrow$ & SMC$\_$SC7 &   47166 \\
SMC$\_$SC3  &213116 & $\leftrightarrow$ & SMC$\_$SC4 &	 22703 &  SMC$\_$SC6  &306569 & $\leftrightarrow$ & SMC$\_$SC7 &   42355 \\
SMC$\_$SC3  &213150 & $\leftrightarrow$ & SMC$\_$SC4 &	 26050 &  SMC$\_$SC6  &306636 & $\leftrightarrow$ & SMC$\_$SC7 &   47253 \\
SMC$\_$SC3  &213173 & $\leftrightarrow$ & SMC$\_$SC4 &	 26121 &  SMC$\_$SC6  &306658 & $\leftrightarrow$ & SMC$\_$SC7 &   47182 \\
SMC$\_$SC3  &217658 & $\leftrightarrow$ & SMC$\_$SC4 &	 29262 &  SMC$\_$SC6  &324260 & $\leftrightarrow$ & SMC$\_$SC7 &   66328 \\
SMC$\_$SC3  &217694 & $\leftrightarrow$ & SMC$\_$SC4 &	 29283 &  SMC$\_$SC6  &324270 & $\leftrightarrow$ & SMC$\_$SC7 &   66246 \\
SMC$\_$SC3  &221608 & $\leftrightarrow$ & SMC$\_$SC4 &	 32510 &  SMC$\_$SC7  &224699 & $\leftrightarrow$ & SMC$\_$SC8 &      42 \\
SMC$\_$SC4  &149900 & $\leftrightarrow$ & SMC$\_$SC5 &	  3339 &  SMC$\_$SC7  &224701 & $\leftrightarrow$ & SMC$\_$SC8 &      19 \\
SMC$\_$SC4  &149963 & $\leftrightarrow$ & SMC$\_$SC5 &	  7187 &  SMC$\_$SC7  &224758 & $\leftrightarrow$ & SMC$\_$SC8 &     118 \\
SMC$\_$SC4  &150078 & $\leftrightarrow$ & SMC$\_$SC5 &	  3346 &  SMC$\_$SC7  &224782 & $\leftrightarrow$ & SMC$\_$SC8 &     139 \\
SMC$\_$SC4  &150110 & $\leftrightarrow$ & SMC$\_$SC5 &	  7164 &  SMC$\_$SC7  &224788 & $\leftrightarrow$ & SMC$\_$SC8 &     146 \\
SMC$\_$SC4  &156285 & $\leftrightarrow$ & SMC$\_$SC5 &	 11508 &  SMC$\_$SC7  &228278 & $\leftrightarrow$ & SMC$\_$SC8 &    3751 \\
SMC$\_$SC4  &160008 & $\leftrightarrow$ & SMC$\_$SC5 &	 16651 &  SMC$\_$SC7  &232170 & $\leftrightarrow$ & SMC$\_$SC8 &    7582 \\
SMC$\_$SC4  &160026 & $\leftrightarrow$ & SMC$\_$SC5 &	 16685 &  SMC$\_$SC7  &232190 & $\leftrightarrow$ & SMC$\_$SC8 &    7594 \\
SMC$\_$SC4  &163514 & $\leftrightarrow$ & SMC$\_$SC5 &	 21054 &  SMC$\_$SC7  &247700 & $\leftrightarrow$ & SMC$\_$SC8 &   22366 \\
SMC$\_$SC4  &163579 & $\leftrightarrow$ & SMC$\_$SC5 &	 21147 &  SMC$\_$SC7  &251622 & $\leftrightarrow$ & SMC$\_$SC8 &   26327 \\
SMC$\_$SC4  &163618 & $\leftrightarrow$ & SMC$\_$SC5 &	 21159 &  SMC$\_$SC7  &259491 & $\leftrightarrow$ & SMC$\_$SC8 &   35110 \\
SMC$\_$SC4  &167178 & $\leftrightarrow$ & SMC$\_$SC5 &	 26510 &  SMC$\_$SC7  &263129 & $\leftrightarrow$ & SMC$\_$SC8 &   38862 \\
SMC$\_$SC4  &167505 & $\leftrightarrow$ & SMC$\_$SC5 &	 26643 &  SMC$\_$SC7  &263135 & $\leftrightarrow$ & SMC$\_$SC8 &   38873 \\
SMC$\_$SC4  &171280 & $\leftrightarrow$ & SMC$\_$SC5 &	 32404 &  SMC$\_$SC7  &266742 & $\leftrightarrow$ & SMC$\_$SC8 &   42511 \\
SMC$\_$SC4  &175348 & $\leftrightarrow$ & SMC$\_$SC5 &	 38527 &  SMC$\_$SC7  &266799 & $\leftrightarrow$ & SMC$\_$SC8 &   42499 \\
SMC$\_$SC4  &179047 & $\leftrightarrow$ & SMC$\_$SC5 &	 43773 &  SMC$\_$SC8  &170177 & $\leftrightarrow$ & SMC$\_$SC9 &      49 \\
SMC$\_$SC4  &186420 & $\leftrightarrow$ & SMC$\_$SC5 &	 54825 &  SMC$\_$SC8  &176830 & $\leftrightarrow$ & SMC$\_$SC9 &    6787 \\
SMC$\_$SC4  &186421 & $\leftrightarrow$ & SMC$\_$SC5 &	 54826 &  SMC$\_$SC8  &179936 & $\leftrightarrow$ & SMC$\_$SC9 &   10101 \\
SMC$\_$SC4  &195793 & $\leftrightarrow$ & SMC$\_$SC5 &	 74747 &  SMC$\_$SC8  &183334 & $\leftrightarrow$ & SMC$\_$SC9 &   13402 \\
SMC$\_$SC4  &198645 & $\leftrightarrow$ & SMC$\_$SC5 &	 74766 &  SMC$\_$SC8  &186699 & $\leftrightarrow$ & SMC$\_$SC9 &   16962 \\
SMC$\_$SC5  &251506 & $\leftrightarrow$ & SMC$\_$SC6 &	   125 &  SMC$\_$SC8  &201625 & $\leftrightarrow$ & SMC$\_$SC9 &   33090 \\
SMC$\_$SC5  &251522 & $\leftrightarrow$ & SMC$\_$SC6 &	    56 &  SMC$\_$SC8  &209987 & $\leftrightarrow$ & SMC$\_$SC9 &   41798 \\
SMC$\_$SC5  &251543 & $\leftrightarrow$ & SMC$\_$SC6 &	   171 &  SMC$\_$SC8  &210034 & $\leftrightarrow$ & SMC$\_$SC9 &   41799 \\
SMC$\_$SC5  &251964 & $\leftrightarrow$ & SMC$\_$SC6 &	  5627 &  SMC$\_$SC9  &144123 & $\leftrightarrow$ & SMC$\_$SC10&    3043 \\
SMC$\_$SC5  &255979 & $\leftrightarrow$ & SMC$\_$SC6 &	  5326 &  SMC$\_$SC9  &144129 & $\leftrightarrow$ & SMC$\_$SC10&    3050 \\
SMC$\_$SC5  &256008 & $\leftrightarrow$ & SMC$\_$SC6 &	  5289 &  SMC$\_$SC9  &146940 & $\leftrightarrow$ & SMC$\_$SC10&    3056 \\
SMC$\_$SC5  &266138 & $\leftrightarrow$ & SMC$\_$SC6 &	 22891 &  SMC$\_$SC9  &147031 & $\leftrightarrow$ & SMC$\_$SC10&    6069 \\
SMC$\_$SC5  &271099 & $\leftrightarrow$ & SMC$\_$SC6 &	 22780 &  SMC$\_$SC9  &147048 & $\leftrightarrow$ & SMC$\_$SC10&    6090 \\
SMC$\_$SC5  &276969 & $\leftrightarrow$ & SMC$\_$SC6 &	 28915 &  SMC$\_$SC9  &152642 & $\leftrightarrow$ & SMC$\_$SC10&    8949 \\
SMC$\_$SC5  &277066 & $\leftrightarrow$ & SMC$\_$SC6 &	 29031 &  SMC$\_$SC9  &155855 & $\leftrightarrow$ & SMC$\_$SC10&   14708 \\
SMC$\_$SC5  &283106 & $\leftrightarrow$ & SMC$\_$SC6 &	 35572 &  SMC$\_$SC9  &166022 & $\leftrightarrow$ & SMC$\_$SC10&   24446 \\
SMC$\_$SC5  &288813 & $\leftrightarrow$ & SMC$\_$SC6 &	 42342 &  SMC$\_$SC9  &175366 & $\leftrightarrow$ & SMC$\_$SC10&   33952 \\
SMC$\_$SC5  &294722 & $\leftrightarrow$ & SMC$\_$SC6 &	 49162 &  SMC$\_$SC10 &112283 & $\leftrightarrow$ & SMC$\_$SC11&    7029 \\
SMC$\_$SC5  &300442 & $\leftrightarrow$ & SMC$\_$SC6 &	 55681 &  SMC$\_$SC10 &114449 & $\leftrightarrow$ & SMC$\_$SC11&    9041 \\
SMC$\_$SC5  &300454 & $\leftrightarrow$ & SMC$\_$SC6 &	 55697 &  SMC$\_$SC10 &116711 & $\leftrightarrow$ & SMC$\_$SC11&   11121 \\
SMC$\_$SC5  &300528 & $\leftrightarrow$ & SMC$\_$SC6 &	 55770 &  SMC$\_$SC10 &122906 & $\leftrightarrow$ & SMC$\_$SC11&   19680 \\
SMC$\_$SC5  &300700 & $\leftrightarrow$ & SMC$\_$SC6 &	 55803 &  SMC$\_$SC10 &124873 & $\leftrightarrow$ & SMC$\_$SC11&   21963 \\
SMC$\_$SC5  &305932 & $\leftrightarrow$ & SMC$\_$SC6 &	 61429 &  SMC$\_$SC10 &134483 & $\leftrightarrow$ & SMC$\_$SC11&   30185 \\
}

\Section{Completeness of the Catalog}

We estimated completeness of the catalog of Cepheid from the SMC in
similar way as for the LMC objects based on comparison of Cepheids
located in overlapping regions of subsequent fields.  10 such regions
exist between our fields  (Fig.~1) allowing to perform 20 tests of
pairing objects from a given and adjacent fields. In total 252 objects
from objects listed in Table~3 should be paired with counterparts in the
overlapping field. We found counterparts in 236 cases which yields the
completeness of our sample equal to 93.7\%.

Taking  into account completeness of detection of stars by the OGLE data
pipeline which was estimated using artificial star tests (Udalski \etal
1998b) and was found to be larger than 99\% for stars as bright as
Cepheids,  the total completeness of our Catalog should be
$\approx92$\%. It is slightly lower than completeness of the LMC catalog
of Cepheids which is a natural consequence of $\approx0.5$~mag fainter
brightness of SMC Cepheids due to larger distance to that galaxy and
larger population of shorter period (fainter) Cepheids in the SMC.

\Section{Discussion}

The OGLE Catalog of Cepheids in the SMC is the largest sample of these
objects detected so far in one galaxy. Together with  smaller sample
presented in the OGLE Catalog of Cepheids from the LMC they constitute  an
unique data set of these stars with high statistical completeness and 
highly homogeneous precise observations --  ideal for analyzing all kinds
of properties of Cepheids including their dependencies on different
properties of stellar environments (\eg on metallicity which is equal
to ${\rm [Fe/H]}=-0.3$~dex, and $-0.7$~dex for the LMC and SMC,
respectively, Luck \etal 1998) etc.

The distribution of Cepheids in the SMC is shown in Fig.~1. Dots
indicate positions of objects within observed fields. The largest
concentration of  Cepheids is found in the central parts of the SMC bar
-- in the fields SMC$\_$SC4--SMC$\_$SC6. Positions of many Cepheids
coincide with areas of star clusters and it is very likely that many of
them are star cluster members. Full list of Cepheids in the SMC clusters
will be presented in a separate paper. Table~5 lists for each field
number of objects from the Catalog, number of all stellar objects
detected  in the field and number of stars brighter than $I_0=17.5$~mag
(approximately the limit of brightness of classical Cepheids in the
SMC).

\MakeTable{lrcr}{12.5cm}{Number of Cepheids and stars in the SMC fields.}
{\hline
\noalign{\vskip3pt}
\multicolumn{1}{c}{Field}&\multicolumn{1}{c}{$N_{Cep}$}& $N_{tot}$ & 
\multicolumn{1}{c}{$N_{I_0<17.5}$}   \\
\noalign{\vskip3pt}
\hline
\noalign{\vskip3pt}
SMC$\_$SC1 &  44~~ & 120002 &  6060 \\
SMC$\_$SC2 &  86~~ & 107326 &  8525 \\
SMC$\_$SC3 & 229~~ & 240045 & 12101 \\
SMC$\_$SC4 & 301~~ & 198201 & 16183 \\
SMC$\_$SC5 & 372~~ & 319850 & 19897 \\
SMC$\_$SC6 & 321~~ & 326367 & 19635 \\
SMC$\_$SC7 & 243~~ & 258006 & 15054 \\
SMC$\_$SC8 & 204~~ & 211115 & 12059 \\
SMC$\_$SC9 & 127~~ & 176832 & 10023 \\
SMC$\_$SC10& 124~~ & 140589 &  9196 \\
SMC$\_$SC11& 117~~ & 120932 &  7611 \\
\hline}

\begin{figure}[p]
\hglue-0.5cm\psfig{figure=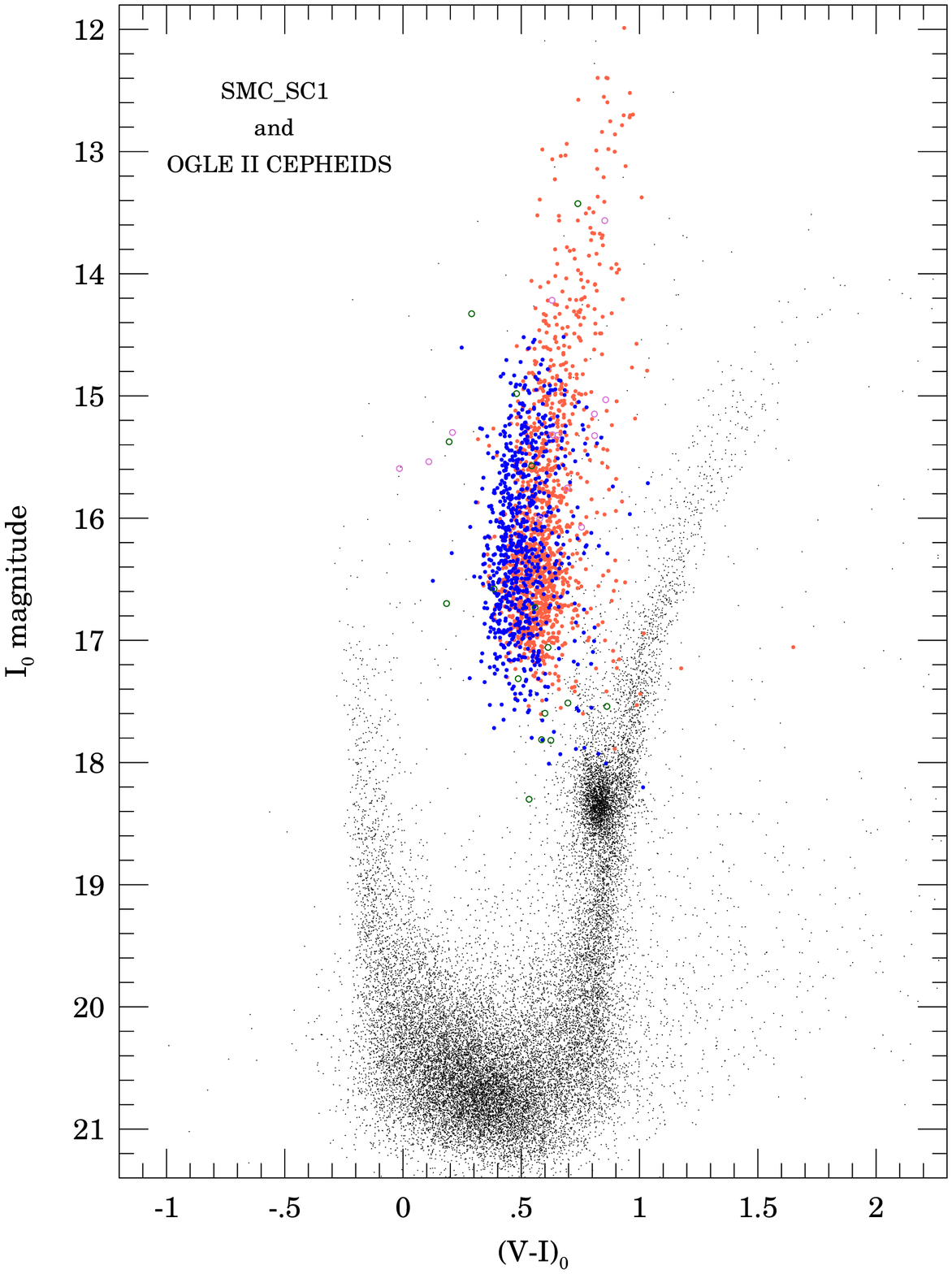,bbllx=35pt,bblly=50pt,bburx=550pt,bbury=730pt,width=13cm,clip=}
\FigCap{Color-magnitude diagram of SMC$\_$SC1 field. Only about 20\% of
field  stars are plotted by tiny dots. Larger dots show positions of FO
and FU classical Cepheids (dark and light dots, respectively). Dark and
light open circles mark positions of objects from the Catalog classified
as FA and BR, respectively. Red clump is a dominating feature at
$I_0\approx18.3$~mag and $(B-V)_0\approx 0.85$~mag.}
\end{figure}

Fig.~4 shows the CMD of the SMC$\_$SC1 field corrected for the mean
reddening in this direction ($E(B-V)=0.07$~mag -- Table~2). Field stars
from this field are plotted by tiny dots. Larger dots indicate positions
of classical Cepheids while open circles positions of the remaining
stars from our Catalog.

\begin{figure}[htb]
\psfig{figure=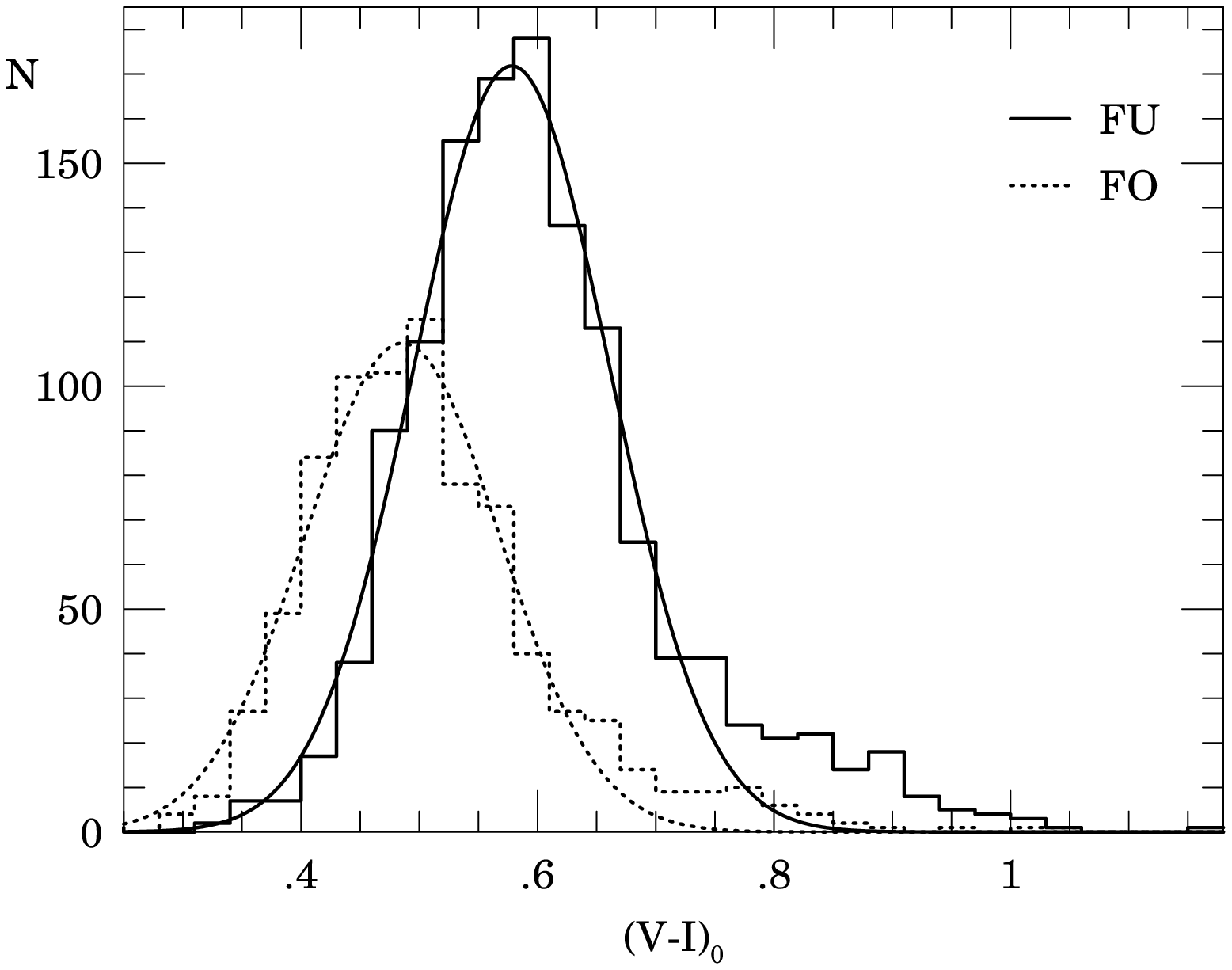,bbllx=50pt,bblly=50pt,bburx=505pt,bbury=410pt,width=12.5cm
,clip=}
\FigCap{Histograms of $(V-I)_0$ color distribution of single-mode 
Cepheids in the SMC. Solid line represents distribution of fundamental
mode pulsators, dotted line -- first overtone objects. The bins are
0.03~mag wide.}
\end{figure}

The vast majority of objects from the Catalog are classical Cepheids
pulsating in the FU and FO modes. Brighter objects (BR) are usually
classical Cepheids, unresolved blends with other star what shifts their
magnitudes and colors and changes shape of the light curve.  Among
fainter objects (FA) one can easily distinguish  Population II Cepheids
which are fainter than classical Cepheids and form a clear sequence
below the $P-L$ relation of classical  FU mode Cepheids (Fig.~2). It is
worth noticing that the population of red giants with light curves
resembling those of pulsating stars -- quite numerous in the LMC, is
practically absent among objects from the SMC.

Fig.~5 presents the distribution of color indices $(V-I)_0$ of classical
FU and FO mode Cepheids. The mode $(V-I)_0$ color and its dispersion are
equal to $(0.577, 0.08)$ and $(0.485, 0.08)$ for  the FU and FO mode
Cepheids in the SMC, respectively. Both average colors are by about
0.02--0.03~mag bluer than those of the LMC sample.

\begin{figure}[htb]
\psfig{figure=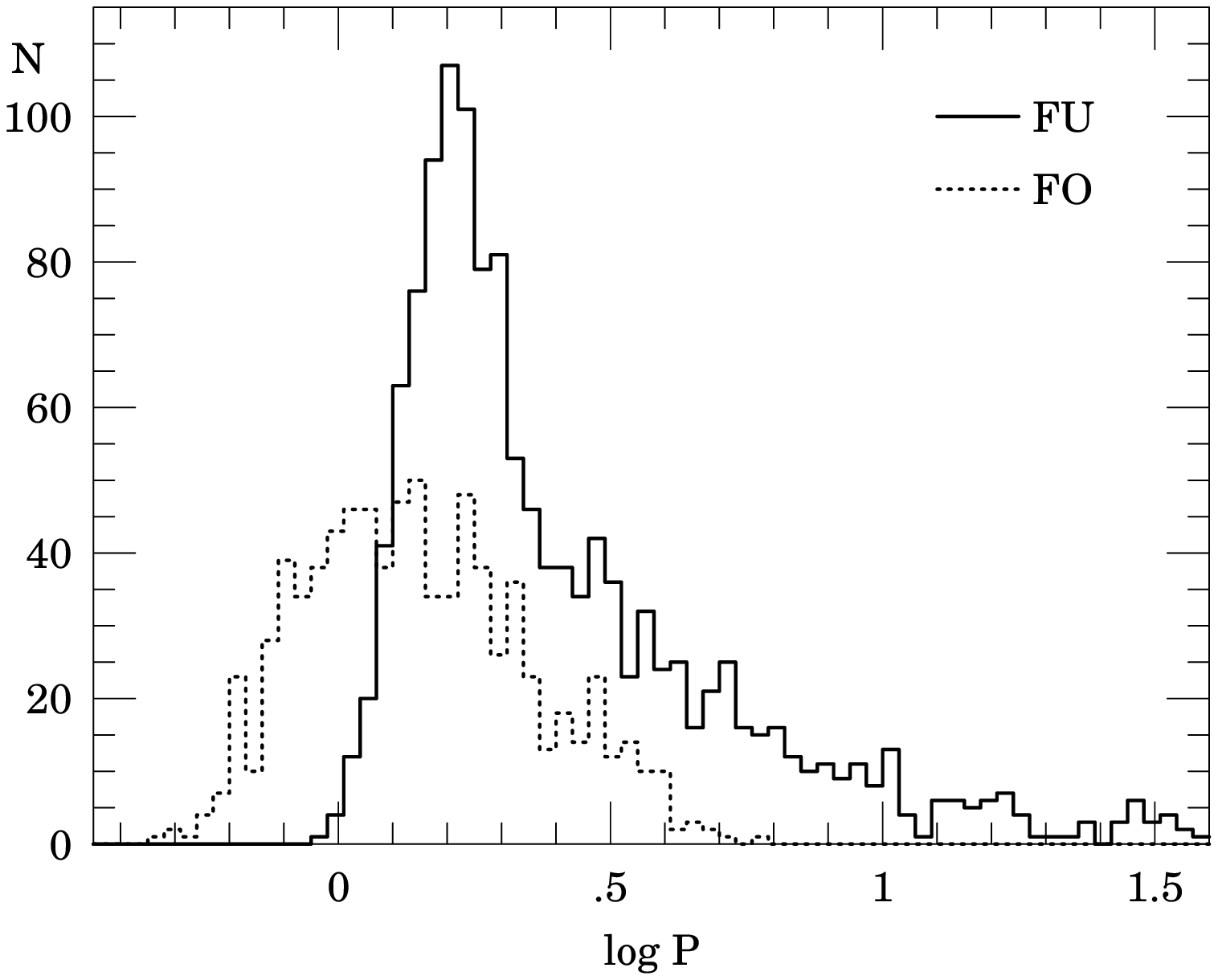,bbllx=50pt,bblly=50pt,bburx=505pt,bbury=410pt,width=12.5cm
,clip=}
\FigCap{Histograms of $\log P$ distribution of single-mode  Cepheids in
the SMC. Solid line represents distribution of fundamental mode
pulsators, dotted line -- first overtone objects. The bins are 0.03
wide in $\log P$.}
\end{figure}

Fig.~6 shows distribution of periods of FU and FO mode classical
Cepheids in the SMC. Mode period of  FU~Cepheids in the SMC is about
1.6~days while for FO objects 1.3~days. The fundamental mode Cepheid
period distribution has a long tail toward long period objects. The
distribution of periods of FO objects is quite different than that
observed in the LMC -- with broad maximum instead of sharp peak as in
the LMC case.

The Catalog of Cepheids from the SMC is available now to the
astronomical community from the OGLE Internet archive:
\begin{center}
{\it http://www.astrouw.edu.pl/\~{}ogle} \\
{\it ftp://sirius.astrouw.edu.pl/ogle/ogle2/var$\_$stars/smc/cep/catalog/}\\
\end{center}
or its US mirror
\begin{center}
{\it http://www.astro.princeton.edu/\~{}ogle}\\
{\it ftp://astro.princeton.edu/ogle/ogle2/var$\_$stars/smc/cep/catalog/}\\
\end{center}

The data include the mean photometry, individual {\it BVI} observations
of all objects and finding charts. Also individual photometry of double
mode and second overtone Cepheids from the SMC (Udalski \etal 1999a,b)
is available there. We plan to update the Catalog in the future when
more observations are collected. We would also appreciate information on
any errors in the Catalog which are unavoidable in so large data set.

\Acknow{We would like to thank Dr.\ B. Paczy\'nski for many discussions
and important suggestions.  The paper was partly supported by  the
Polish KBN grants 2P03D00814 to  A.~Udalski and 2P03D00916 to
M.~Szyma{\'n}ski. Partial support for the OGLE  project was provided
with the NSF  grant AST-9820314 to B.~Paczy\'nski. We  acknowledge usage
of The Digitized Sky Survey which was produced at the Space  Telescope
Science Institute based on photographic data obtained using The UK 
Schmidt Telescope, operated by the Royal Observatory Edinburgh.}

\end{document}